\documentclass[prl, reprint, superscriptaddress]{revtex4-2}
\usepackage{newtxmath}
\usepackage{newtxtext}
\usepackage{dcolumn}
\usepackage{braket}
\usepackage{graphicx}
\usepackage{orcidlink}
\usepackage{color}
\definecolor{bblue}{rgb}{0, 0.0, 0.8}
\definecolor{rred}{rgb}{0.7, 0.0, 0.0}

\usepackage{hyperref}
\hypersetup{
    colorlinks=true,        
    linkcolor=rred,          
    citecolor=rred,        
    filecolor=rred,      
    urlcolor=rred           
}

\usepackage{dcolumn}

\begin{document}
\title{Efficiency in a repetitive pulse magnet}
\author{Akihiko Ikeda \orcidlink{0000-0001-7642-0042}}
\email[email: ]{a-ikeda@uec.ac.jp}
\affiliation{Department of Engineering Science, University of Electro-Communications, Chofu, Tokyo 182-8585, Japan}

\author{Yuto Ishii \orcidlink{0000-0001-6849-8915}}
\author{Yasuhiro H. Matsuda\orcidlink{0000-0002-7450-0791}}
\author{Go Yumoto \orcidlink{0000-0002-6489-343X}}
\author{Ayumi Abe}
\author{Ryusuke Matsunaga \orcidlink{0000-0001-9674-5100}}
\affiliation{Institute for Solid State Physics, University of Tokyo, Kashiwa, Chiba 277-8581, Japan}

\date{\today}

\begin{abstract}
A repetitive-pulse magnet is a promising tool when combined with repetitive excitations, such as pulsed lasers.
Technically, the repetition and the magnetic field values in a repetitive-pulse magnet are limited by the Joule heating in the coil.
Here, we analytically examine the relationship between the coil's dimensions and its efficiency, assuming negligible heating of the coil, to design an optimized high-repetition, high-magnetic-field coil.
We calculated the dependence of the maximum magnetic field, energy loss, pulse duration, form factor, impedance, and maximum current on the coil's geometry.
We found that the smaller the coil, the more pulses and the more intense the magnetic fields we can obtain under a given condition.
We argue that the obtained trend arises from a complex interplay among various parameters.
\end{abstract}

\maketitle
\section{Introduction}

Repetitive-pulse magnets \cite{Jiang2020RSI, Jiang2021RSI, Zhang2024IEEE} have been used to enhance the signal-to-noise ratio in experiments such as Muon spin spectroscopy, neutron scattering, and other beam-based studies \cite{MotokawaPhysicaB1989, MotokawaIEEE1996, InadaPRL2017}.
Integrating these magnets with synchronized femtosecond pulsed lasers is of particular importance [Fig. \ref{fig01}(a)]. 
For instance, single-shot spectroscopy techniques synchronized with pulsed magnets have enabled terahertz time-domain spectroscopy under high magnetic fields \cite{NoeOE2016, LiPRB2019}. 
Furthermore, pump-probe spectroscopy under high magnetic fields is also in demand for investigating nonequilibrium dynamics \cite{LMittendorffNP2015, YumotoPRL2018, MatsudaPRL2023}. 
However, such studies require a high repetition rate to accommodate the wide range of parameters that must be varied, including pump-probe delay, pump intensity, pump and/or probe polarization, and sample temperature.   
The primary technical challenge in developing high-repetition-rate pulsed magnets is coil heating.
We need to consider how to minimize coil heating while maximizing magnetic field intensity and repetition rate.
A successful coil operation exhibits a small temperature rise following each pulsed magnetic-field generation, so that we can keep the coil temperature within the available cooling power.

Here, we analyze the relationships among the dimensions of the pulse magnet, energy dissipation, and magnetic field strength under given conditions, such as the charging voltage and capacitance, assuming negligible coil heating and constant circuit resistivity.
The formalism relies on Ref. \cite{Knoepfel, Champion, Grover}, which is reformulated here for the present purpose.
We are interested in a high-repetition-rate pulse magnet capable of more than 1 pulse per second and a magnetic field greater than 1 T, to be combined with repetitive excitations such as pulsed lasers.

\section{Model}

\subsection{Energy in a pulse magnet}
\begin{figure}
\begin{center}
\includegraphics[width = \columnwidth]{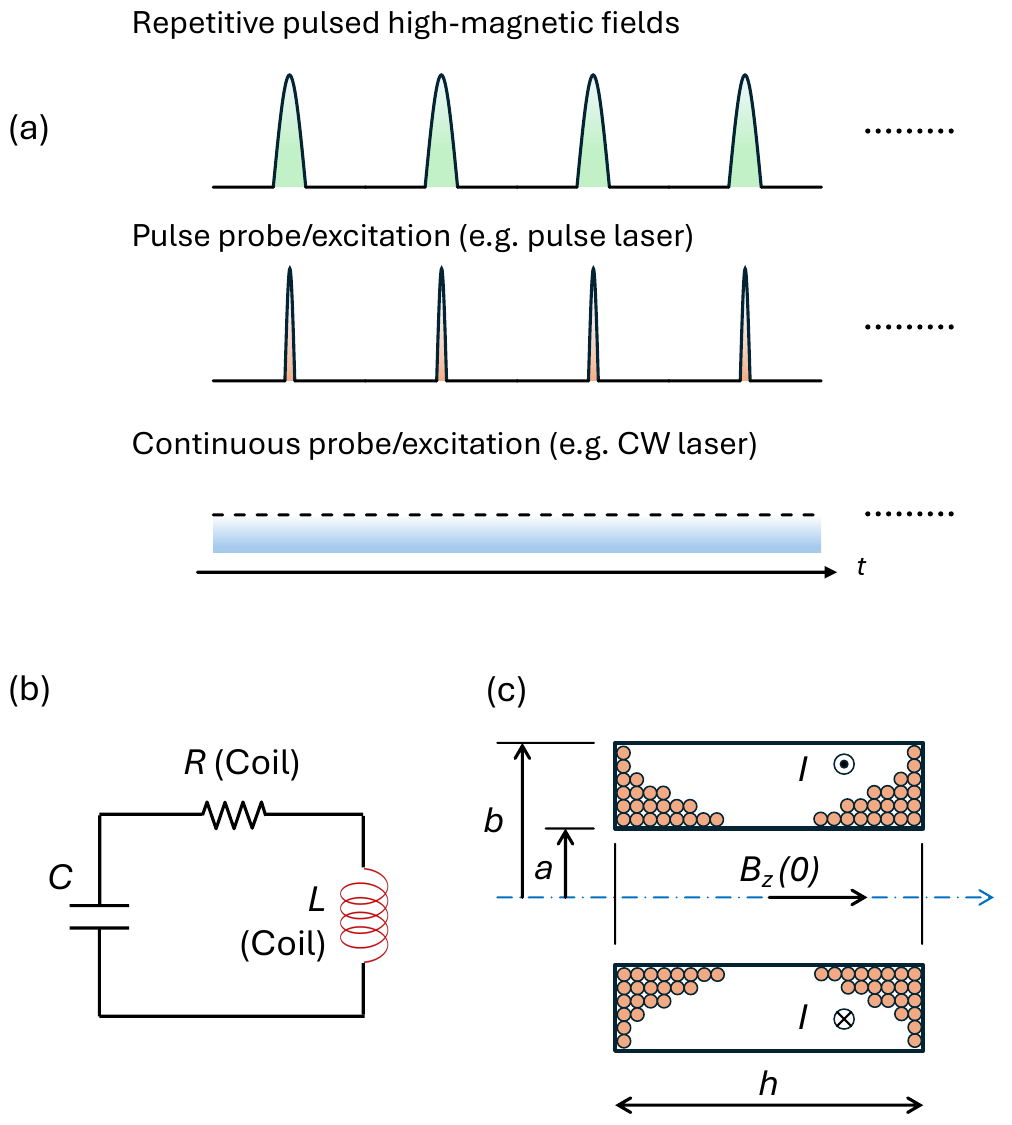}
\caption{
(a) Schematic of the experiment combining a repetitive pulsed magnetic field and a pulsed excitation source or a continuous excitation source.
(b) Electric circuit, and (c) Dimensions of coil.
\label{fig01}}
\end{center}
\end{figure}

We consider the free discharge in an electric circuit composed of a capacitor and a coil.
The circuit is represented in Fig. \ref{fig01}(b).
A solenoid coil with the inner radius $a$ and outer radius $b$ and the axial height $h$ as represented in Fig. \ref{fig01}(c).
The energy conservation is described as 
\begin{equation}
E_{0} = E_{C} + E_{L} + E_{\rm{loss}},
\end{equation}
where $E_{0}$, $E_{C}$, $E_{L}$, and $E_{\rm{loss}}$ are the total energy defined as $CV^{2}/2$ with the capacitance $C$ and the initial charging voltage $V_{0}$, the energy stored at the capacitor, the energy stored at the inductor, and the energy dissipated at the resistor, respectively. 
In the ideal limit, the uniform magnetic field is confined within the solenoid coil's bore.
In this case,
\begin{align}
E_0 = \frac{\pi a^2 h B_{\rm{m0}}^{2}}{2\mu_{0}}\\
B_{\rm{m0}} = \sqrt{\frac{2\mu_{0}E_0}{\pi a^2 h}}\label{bm0}
\end{align}
Due to the form factor $S$ and the dissipation factor $D$, the actual magnetic field at the center of the coil is smaller than $B_{\rm{m0}}$ as
\begin{equation}
B_{\rm{m}} = SDB_{\rm{m0}} = SD\sqrt{\frac{2\mu_{0}E_0}{\pi a^2 h}}\label{sdbm0}
\end{equation}
where $S$ is a function of the dimensions of the coil ($a$, $b$, $h$) as shown in Eq. \ref{fig01}.
$S$ and $D$ take values between unity and 0.

\subsection{Discharge circuit}

\begin{figure}
\begin{center}
\includegraphics[width = \columnwidth]{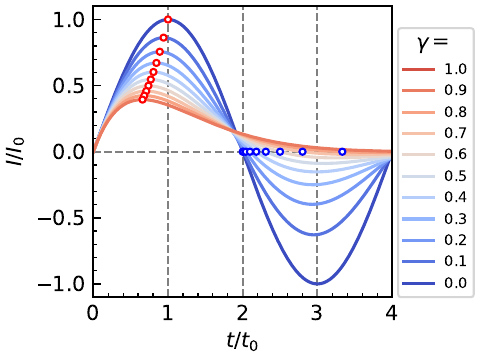}
\caption{
Normalized pulsed currents $I/I_{0}$ as a function of the normalized time $t/t_0$ for the values of $\gamma$ in the range $0\leq\gamma\leq1$.
\label{iprofile}}
\end{center}
\end{figure}

The circuit equation is shown as
\begin{equation}
\frac{1}{C}\int I dt + L\frac{dI}{dt} + RI = 0,
\end{equation}
where
\begin{equation}
I = \frac{dq}{dt}.
\end{equation}
The initial boundary conditions $t=0$ $I = 0$ and $q = q_0$, with $ q_0 = CV_{0}$, the solutions are known to be classified into the four cases, $\gamma =0$ (Undamped), $0 <\gamma <1$ (Underdamped), $\gamma =1$ (Critically damped), $\gamma > 1$ (Overdamped)
, where the dissipation ratio $\gamma$ is described as 
\begin{equation}
\gamma = \frac{R}{2}\sqrt{\frac{C}{L}}.\label{def_gamma}
\end{equation}
Here, we focus on the underdamped case.
The solution is 
\begin{equation}
I = \frac{V_{0}}{\omega L} e^{-at}\sin\omega t, \label{do}
\end{equation}
with $a = R/(2L)$, and the angular frequency $\omega$  
\begin{equation}
\omega = \sqrt{\frac{1}{LC}-\frac{R^{2}}{4L^{2}}} = \omega_{0}\sqrt{1-\gamma^{2}},
\end{equation}
where the dissipationless angular frequency $\omega_{0} = 1/\sqrt{LC}$. 
Eq. \ref{do} is also described as
\begin{equation}
\frac{I}{I_{0}} = \frac{1}{\sqrt{1-\gamma^{2}}}\exp\left( -\gamma \frac{\pi}{2}\frac{t}{t_{0}}\right)\sin \left(\sqrt{1-\gamma^{2}}\frac{\pi}{2}\frac{t}{t_{0}}\right), \label{di}
\end{equation}
using the dissipationless current amplitude $I_{0}$, the dissipationless quarter duration $t_{0}$ described as 
\begin{align}
I_{0}&=\frac{V_{0}}{\omega_{0} L}, \\
t_{0} &= \frac{\pi}{2\omega_{0}}.
\end{align}
Eq. \ref{di} shows $\gamma$ dependence of the normalized current $I/I_{0}$ profile as a function of normalized time $t/t_0$ as shown in Fig. \ref{iprofile}.

Here, we consider the time required to reach the current maximum, $t_{\rm{m}}$.
Derivative of Eq. \ref{do} is
\begin{align}
\frac{dI}{dt} =  \frac{V_{0}}{ L} \sqrt{\frac{1}{1-\gamma^{2}}} e^{-at}\sin \left(\omega t + \alpha\right).
\end{align}
Clearly, the condition $\omega t_{\rm{m}} + \alpha = \pi$ satisfies $\left. \frac{dI}{dt}\right|_{t = t_{\rm{m}}} = 0$.
Hence, one obtains $t_{\rm{m}}$ as 
\begin{equation}
t_{\rm{m}} =\frac{1}{\omega} \left(\pi - \alpha \right)=\frac{1}{\omega} \arcsin\left(\sqrt{1-\gamma^{2}}\right),
\end{equation}
and with normalization one obtaines a form that dependes only on $\gamma$ as 
\begin{equation}
\frac{t_{\rm{m}}}{t_{0}} =\frac{2}{\pi}\frac{1}{\sqrt{1-\gamma^{2}}} \arcsin\left(\sqrt{1-\gamma^{2}}\right).
\end{equation}
Then, the maximum current is obtained using $t_{\rm{m}}$ and Eq. \ref{do} with normalization as 
\begin{align}
\frac{I_{\rm{m}}}{I_{0}} &=\frac{1}{\sqrt{1-\gamma^{2}}}e^{-at_{\rm{m}}}\sin(\omega t_{\rm{m}})\\
 &=\exp\left(-at_{\rm{m}}\right)\\
 &=\exp\left(-\frac{\gamma}{\sqrt{1-\gamma^{2}}}\arcsin\left(\sqrt{1-\gamma^{2}}\right) \right),\label{im0}
\end{align}
which depends only on $\gamma$.
Lastly, we consider the time to reach the zero current $t_{z}$ as 
\begin{equation}
t_{z} = \frac{\pi}{\omega} =\frac{\pi}{\omega_{0}}\frac{1}{\sqrt{1-\gamma^{2}}}.
\end{equation}
With normalization, we obtain
\begin{equation}
\frac{t_{z}}{t_{0}} = \frac{2}{\sqrt{1-\gamma^{2}}}, 
\end{equation}
which again depends only on $\gamma$.

\begin{figure}
\begin{center}
\includegraphics[width = \columnwidth]{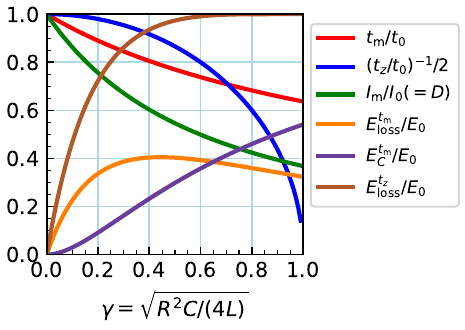}
\caption{
The normalized values of $t_{\rm{m}}/t_{0}$, $(t_{z}/t_{0})^{-1}/2$, $I_{\rm{m}}/t_{0}$, $E_{\rm{loss}}^{t_{\rm{m}}}$, $E_{\rm{C}}^{t_{\rm{m}}}$, and $E_{\rm{loss}}^{t_{z} }$ shown as a function of $\gamma$ from 0 to unity.
\label{gammas}}
\end{center}
\end{figure}

Fig. \ref{gammas} shows the $\gamma$ dependence of the normalized values,  $t_{\rm{m}}/t_{0}$, $I_{\rm{m}}/I_{0}$, and $(t_{z}/t_{0})^{-1}/2$.
All these values exhibit monotonic decreases from 1 as $\gamma$ increases from 0 to 1.
The behavior is also apparent from Fig. \ref{iprofile}, where the red and blue circles indicate the coordinate ($t_{\rm{m}}/t_{0}$, $I_{\rm{m}}/I_{0}$) and the coordinate ($t_{z}/t_{0}$, 0), respectively, as a function of $\gamma$.

\subsection{Energy dissipation}
To account for energy dissipation during pulsed magnetic field generation, we calculate the Joule heating energy from $t = 0$.
First, we calculate the Joule heating up to the time $t_{\rm{m}}$ using Eq. \ref{do}.
We obtain 
\begin{align}
E_{\rm{loss}}^{t_{\rm{m}}}&= \int_{0}^{t_{\rm{m}}} RI^{2} dt\\
&= \frac{CV_{0}^{2}}{2} \left[1 -e^{-2at_{\rm{m}}}\left(1 + 4\gamma^{2}\right)\right]. \label{elossm}
\end{align}
We normalize Eq. \ref{elossm} using $E_{0}$ as 
\begin{equation}
\frac{E_{\rm{loss}}^{t_{\rm{m}}}}{E_{0}}=  1 -\left(\frac{I_{\rm{m}}}{I_{0}}\right) ^{2}\left(1 + 4\gamma^{2}\right).\label{elossm0}
\end{equation}
which depends on $\gamma$ and $I_{\rm{m}}/I_{0}$ which depends only on $\gamma$ (Eq. \ref{im0}).
Thus, Eq. \ref{elossm0} depends only on $\gamma$.

At $t = t_{\rm{m}}$,  $V_{L} = 0$.
Hence, $V_{C}= -RI_{\rm{m}}$.
We can calculate the remaining energy at the capacitance as 
\begin{equation}
E_{C} = \frac{1}{2}C(RI_{\rm{m}})^{2}
\end{equation}
and the normalized form as 
\begin{equation}
\frac{E_{C}}{E_0}=4\gamma^{2}\left(\frac{I_{\rm{m}}}{I_0}\right)^{2},\label{eceo}
\end{equation}
which depends only on $\gamma$.

With Eq. \ref{elossm0} and Eq. \ref{eceo}, we can calculate $D$ as
\begin{align}
D &= \sqrt{\frac{E_{L}}{E_{0}}} = \sqrt{\frac{E_{0}- E_{\rm{loss}}^{t_{\rm{m}}}- E_C}{E_{0}}} =  \frac{I_{\rm{m}}}{I_{0}},
\label{DD}
\end{align}
which is equal to Eq. \ref{im0} dependent only on $\gamma$.

Next, we calculate the dissipated energy from $t=0$ to $t_{z}$.
Similarly,
\begin{align}
E_{\rm{loss}}^{t_{z}}
&= \int_{0}^{t_{z} } RI^{2} dt\\
&= \frac{CV_{0}^{2}}{2} \left[ 1 - \exp\left(- \frac{2\pi}{\sqrt{\frac{1}{\gamma^{2}} - 1}}\right) \right].
\end{align}
Further, the normalized form is obtained as 
\begin{equation}
\frac{E_{\rm{loss}}^{t_{z}}}{E_{0}}=  1 - \exp\left(- 2\pi\frac{\gamma}{\sqrt{1 - \gamma^{2}}}\right),
\end{equation}
which depends only on $\gamma$.
The obtained values $E_{\rm{loss}}^{t_{\rm{m}}}/E_{0}$, $E_{C}/E_0$,  and $E_{\rm{loss}}^{t_{z}}/E_{0}$ are plotted in Fig. \ref{gammas} as a function of $\gamma$.

\subsection{$\gamma$ and a coil}
So far, we have expressed all important parameters using $\gamma$.
Here, we show that $\gamma$ can be expressed in terms of the coil dimensions and the capacitance.
First, we express the inductance and resistance of the coil as 
\begin{align}
L_{L} &= \mu_{0}N^{2}\frac{\pi a}{h}K_{\rm{L}}, \label{LL}\\
R_{L} &= \frac{\pi }{\sigma}\frac{N^{2}}{fh}\frac{b+a}{b-a}. \label{RL}
\end{align}
Using these expressions and Eq. \ref{def_gamma}, $\gamma$ can be expressed in terms of the coil's dimension and $C$, if we neglect the circuit's residual impedance, as 
\begin{equation}
\gamma  = \frac{N}{2\sigma f} \frac{b+a}{b-a}\sqrt{ \frac{\pi C}{\mu_{0}K_{\rm{L}}ha^{2}}}, 
\end{equation}
where $N$, $\sigma$, and $f$ are the total winding number, electrical conductivity, and the filling factor of magnet wire, respectively. Total winding number is expressed as $N  = f (b-a ) h / \Sigma$, using wire cross section $\Sigma$.

\section{Results and Discussion}
We now calculate the following parameters $B_{\rm{m}}$, $B_{\rm{m0}}$, $S$, $D$, $R_{L}$, $L_{L}$, $I_{\rm{m}}$, $I_{0}$, Magnet efficiency, $t_{\rm{m}}$, $t_{z}$, $E_{\rm{loss}}^{t_{\rm{m}}}$, $E_{\rm{loss}}^{t_{\rm{m}}}/E_{0}$, $E_{\rm{loss}}^{t_{z}}$, $E_{\rm{loss}}^{t_{z}}/E_{0}$, $B/I$, using given coil's dimension and $C$, $V_{0}$. 
We plotted the calculated parameters on $b-h$ plane with the conditions of $a=3$ mm, $C=500$ $\mu$F, and $V_{0} = 150$ V, and a circular magnet wire with diameter of $\phi1$ mm  ($f = 0.7$).

\begin{figure*}
\begin{center}
\includegraphics[width = \textwidth]{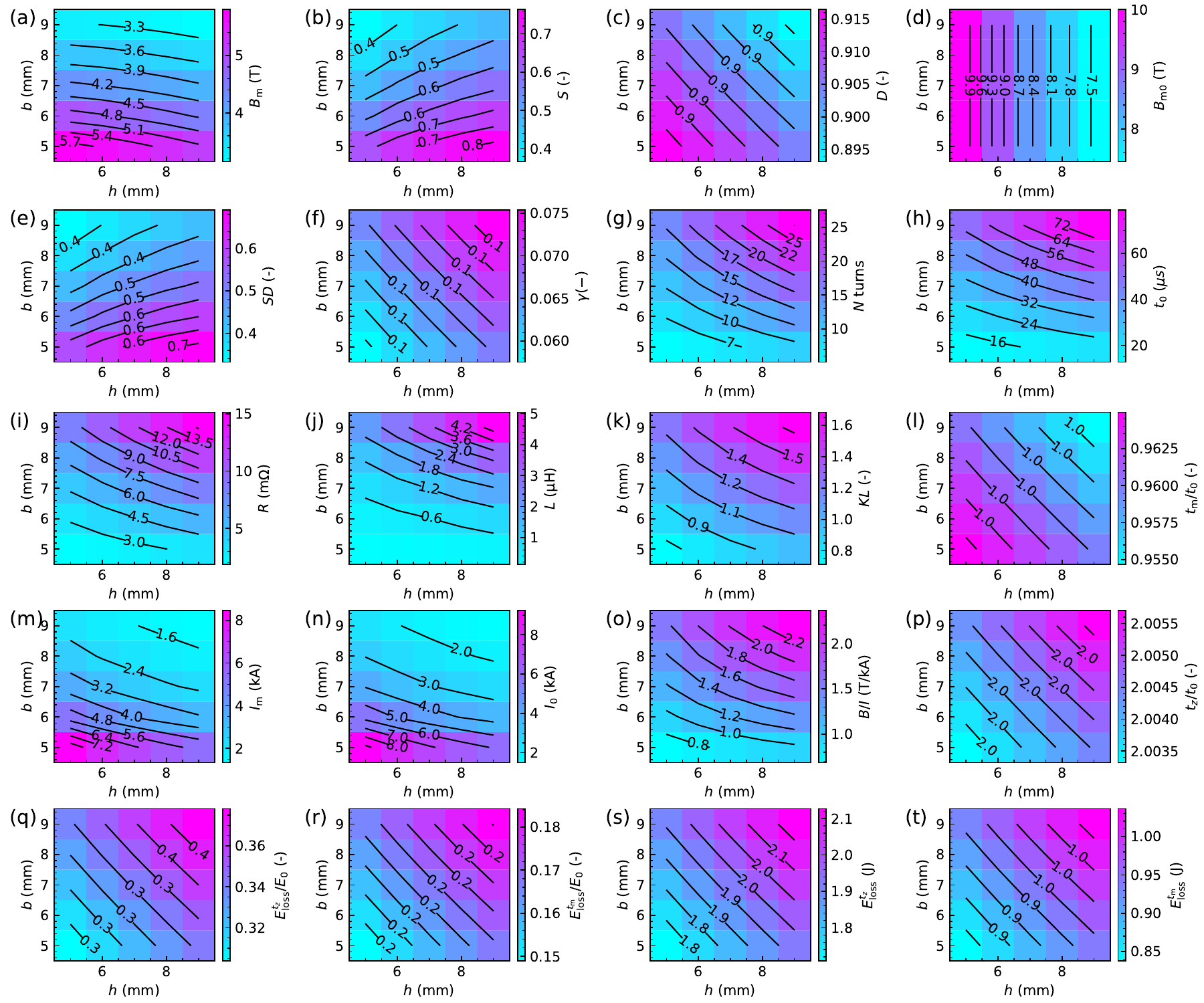}
\caption{
$B_{\rm{m}}$, $B_{\rm{m0}}$, $S$, $D$, $R_{L}$, $L_{L}$, $I_{\rm{m}}$, $I_{0}$, $B/I$, $t_{\rm{m}}$, $t_{z}$, $E_{\rm{loss}}^{t_{\rm{m}}}$, $E_{\rm{loss}}^{t_{\rm{m}}}/E_{0}$, $E_{\rm{loss}}^{t_{z}}$, $E_{\rm{loss}}^{t_{z}}/E_{0}$, $B/I$ calculated assuming $a=3$ mm, $C=500$ $\mu$F, $V_{0} = 150$ V, with a circular magnet wire with diameter of $\phi1$ mm.
\label{results}}
\end{center}
\end{figure*}

It is highlighted that we found that the highest magnetic field $B_{\rm{m}}$ and the smallest energy dissipation $E_{\rm{loss}}^{t_{z}}/E_{0}$  are simultaneously realized using smaller $b$ and $h$ as indicated in Fig. \ref{results}(a) and  \ref{results}(s), respectively.
Therefore, a prompt conclusion is that for a high-repetition and high-magnetic-field magnet, we should employ smaller $b$ and $h$.
The result is simple; however, it results from the competing factors, which is not straightforward.
We discuss this competition below.

\subsection{Anatomy of $B_{\rm{m}}$ }

$B_{\rm{m}}$ is the product of $S$, $D$, and $B_{\rm{m0}}$ as shown in Eq. \ref{sdbm0}.
The tendency of $B_{\rm{m}}$ is rooted in $S$, $D$, and $B_{\rm{m0}}$.
In Fig. \ref{results}(b), $S$ has a larger value with larger $h$ and smaller $b$, which is a different trend from that of $B_{\rm{m}}$ with respect to $h$ in Fig. \ref{results}(a).
$S$ is a form factor that takes unity with the ideal solenoid coil, where the thinnest coil thickness $b-a$ and the longest coil axis $h$ result in the perfect concentration of the uniform magnetic field inside the coil.
In Fig. \ref{results}(c), $D$ shows a similar trend to that of $B_{\rm{m}}$, although the variation of the value is small.
In fact, $SD$ is shown in Fig. \ref{results}(e) showing that $SD$ is governed by $S$ rather than $D$.
$D(=I_{\rm{m}}/I_{0})$ is a function of $\gamma$ [Eqs.\ref{DD} and \ref{im0}], whose relation is shown in Fig. \ref{gammas}.
In Fig. \ref{results}(f), $\gamma$ is smaller with smaller $b$ and $h$.
The energy dissipation is smallest when $\gamma$ is close to 0.
The value of $\gamma$ is between 0.11 and 0.15, which is close enough to 0.
This is the reason why the variation of $D$ is small in Fig. \ref{results}(c).
$B_{\rm{m}0}$ shows the magnetic field value in the case that all the energy is converted and confined inside the coil, which is thus dependent on $h$.
As a result of multiplication of $SD$ and $B_{\rm{m}0}$, we obtain $B_{\rm{m}}$.
It is now clear that it results from competing parameters, and the outcome may change due to a delicate balance.

\subsection{Anatomy of energy dissipation}

The limiting factor of the repetition and $B_{\rm{m0}}$ in the repetitive pulse magnet is the energy dissipation at the coil by Joule heating.
In order to make our pulse more repetitive, $E_{\rm{loss}}^{t_{z}}$, the energy dissipation per pulse at the coil, should be minimized.
Fig. \ref{results}(s) shows that $E_{\rm{loss}}^{t_{z}}$ has a tendency that they are minimum when $b$ and $h$ are smaller.
It is somewhat counterintuitive that a higher magnetic field and the lowest energy dissipation can be achieved simultaneously.
To understand this, we recall that the Joule heating component is $RI^{2}\Delta t$. 

First, we see the pulse duration, $t_{z}$.
$t_0$, $t_{\rm{m}}/t_0$, and $t_{z}/t_0$ are shown in Figs. \ref{results}(h), \ref{results}(l), \ref{results}(p), where a large variation of $t_0$ that decreases with decreasing $b$ and $h$ is apparent with small variation of $t_{\rm{m}}$ and $t_{z}$.
The small energy dissipation is obtained at small values of $b$ and $h$ because they exhibit a short pulse duration, which should lead to the small Joule heating, $RI^{2}\Delta t$.
For $R$, see Fig. \ref{results}(i), where one sees a significantly smaller $R$ with smaller $b$ and $h$, which is reasonable because $R$ is proportional to the total length of the wire, which is smaller with smaller $b$ and $h$.
For $I_{\rm{m}}$ and $I_{0}$, see Figs. \ref{results}(m) and Figs. \ref{results}(n), where one sees a significantly larger $I$ value with smaller $b$ and $h$, which is a counter factor in the Joule heating.
Now, one sees that  $E_{\rm{loss}}^{t_{z}}$ is the result of the competition between $R$, $\Delta$, and $I$, where the former two and the last one have contrary trends.

Similary to $E_{\rm{loss}}^{t_{z}}$, $E_{\rm{loss}}^{t_{\rm{m}}}$, the energy dissipation at the coil by the time $t_{\rm{m}}$,  has the same tendency as shown in Fig. \ref{results}(t).
Fig. \ref{results}(q) and (r) shows $E_{\rm{loss}}^{t_{z}}/E_{0}$ and $E_{\rm{loss}}^{t_{\rm{m}}}/E_{0}$, where the normalization factor is the total energy $E_{0}=CV^{2}/2$ which is a constant value.

\subsection{$B$ and $I$}
the conversion rate $B/I$ is an important parameter, which is shown in Fig. \ref{results}(o).
One sees that a smaller $B/I$ is obtained using smaller $b$ and $h$.
This is due to the smaller winding number $N$ at smaller $b$ and $h$ as shown in Fig. \ref{results}(g).
The contrary trend in $B_{\rm{m}}$ is a result of the compensation by the large value of $I_{\rm{m}}$ at smaller $b$ and $h$.
The large $I_{\rm{m}}$ is realized by the short pulse duration $t_{\rm{m}}$ and $t_{z}$, which is a result of small $L$ as shown in Fig. \ref{results}(j). 
Intuitively, a small $L$ would lead to a small $B_{\rm{m}}$, which is contrary to the present result.

\subsection{Effect of residual $R$ and $L$}
We neglected residual $R$ and $L$ in the present discussion.
Without residual $R$ and $L$, the highest magnetic field is efficiently obtained by using the smallest number of windings (i.e., Single-turn coil).
However, practical pulse power systems have residual impedance due to components such as the thyristor switch and the capacitor.
The limiting factor in practical cases will be the residual impedance, where one needs to pay efforts to minimize it.

\subsection{Relation to the single-shot pulsed high magnetic fields}
One obtains the highest magnetic field with fewer turns in the present calculation.
This is evidenced by the single-turn coil method, where only a single turn of a coil generates over 100 T and up to 300 T \cite{HerlachRPP1999, IkedaAPL2022, IkedaPRL2025}.
The total current amounts to 1 MA, where not only the heating of the coil but also the pulsed magnet's explosion due to magnetic-field pressure dictates the methodology.

A more intense magnetic field beyond 1000 T is obtained in the flux compression method \cite{NakamuraRSI2018_002}, where a more important factor is the realization of the high-energy-density state using a high pressure, represented as $B^{2}/(2\mu_{0})$ [J/m$^{3}$ (=Pa)].

Even below 100 T, the single-shot pulsed magnetic field provides a higher magnetic field with a smaller footprint than a static electromagnet.
The main motivation of the pulsed operation is to avoid the coil's heating.
Various pulse magnets have been devised to achieve a high magnetic field beyond 40 T in a single shot, including nondestructive magnets up to 80 T \cite{MiyataIEEE2026} and a miniature coil up to 40 T \cite{IkedaPRR2020, IkedaJAP2024, NodaAPL2025, IkedaPRR2026}.
Such single-shot pulse magnets are used widely in research areas, including condensed matter, plasma physics, and fundamental physics \cite{BattestiPR2018}.

\section{Summary}
To achieve high energy efficiency and high magnetic fields, we considered various competing factors in the design of a repetitive pulsed magnet.
We conducted analytical calculations of various parameters in a repetitive-pulse magnet, and the results are then plotted for a specific case.
It is concluded that one should decrease the coil length and the coil thickness in a given condition.
The present study demonstrates a delicate balance among the parameters of the coil design to understand the obtained trend.

\begin{acknowledgments}
This work is supported by the JST FOREST program No. JPMJFR222W and No. JPMJFR2240, JSPS Grant-in-Aid for Scientific Research on Innovative Areas (A) (1000 T Science) 23H04861, 23H04859, (Chimera Quasiparticle) 25H02118, (Correlation Design Science) 26H01296 and Grant-in-Aid for Scientific Research (B) 24K00550, 23H01121, Scientific Research (C) 26K08143, Early-Career Scientists 24K17003, and MEXT LEADER program No. JPMXS0320210021.
\end{acknowledgments}

\bibliography{coil}
\end{document}